\documentclass[superscriptaddress]{revtex4}
\usepackage{graphicx}
 \usepackage{color}
 \usepackage{bm}
 \usepackage[sort&compress]{natbib}
 \usepackage{mathrsfs} 
 \usepackage{amsmath,lscape,epsfig}
\usepackage[normalem]{ulem}

\begin{document}
%%%%%%%%%%%%%%%%%%%%%%%%%%%%%%%%%%%%%%%%%%%%%%%%%%%%%%%%%%%%%%%%%%%%%%%%%%%%%%%%
\title{Comparative study of three-nucleon force models in nuclear matter}
%%%%%%%%%%%%%%%%%%%%%%%%%%%%%%%%%%%%%%%%%%%%%%%%%%%%%%%%%%%%%%%%%%%%%%%%%%%%%%%%%
\author{Domenico Logoteta}
\affiliation{INFN, Sezione di Pisa, Largo Bruno Pontecorvo 3, I-56127 Pisa, Italy} 
\author{Isaac Vida\~na}
\affiliation{Centro de F\'{\i}sica Computacional, Department of Physics,
                  University of Coimbra, 3004-516 Coimbra, Portugal}
\author{Ignazio Bombaci}
\affiliation{Dipartimento di Fisica, Universit\'a di Pisa, Largo Bruno Pontecorvo 3, I-56127 Pisa, Italy }
\affiliation{INFN, Sezione di Pisa, Largo Bruno Pontecorvo 3, I-56127 Pisa, Italy}
\author{Alejandro Kievsky}
\affiliation{INFN, Sezione di Pisa, Largo Bruno Pontecorvo 3, I-56127 Pisa, Italy}
%%%%%%%%%%%%%%%%%%%%%%%%%%%%%%%%%%%%%%%%%%%%%%%%%%%%%%%%%%%%%%%%%%%%%%%%%%%%%%%%%%%%%%%%%%%
\newcommand {\bra}{\langle}
\newcommand {\ket}{\rangle} 
\newcommand {\rv}{{\bf r}}
\newcommand {\tv}{{\bf t}}
\newcommand {\sv}{{\bf s}}
\newcommand {\yv}{{\bf y}}
\newcommand {\xv}{{\bf x}}
\newcommand {\xiv}{{\bf \xi}}
\newcommand {\etav}{{\bf \eta}}

%%%%%%%%%%%%%%%%%%%%%%%%%%%%%%%%%%%%%%%%%%%%%%%%%%%%%%%
\begin{abstract}

We calculate the energy per particle of symmetric nuclear matter and pure neutron matter 
using the microscopic many-body Brueckner-Hartree-Fock (BHF) approach and employing the Argonne V18 (AV18)
nucleon-nucleon (NN) potential supplemented with two different three-nucleon force models 
recently constructed to reproduce the binding energy of $^3$H, $^3$He and $^4$He nuclei as well as 
the neutron-deuteron doublet scattering length. 
We find that none of these new three-nucleon force models is able to reproduce simultaneously the empirical saturation 
point of symmetric nuclear matter and the properties of three- and four-nucleon systems. 
\end{abstract}
%%%%%%%%%%%%%%%%%%%%%%%%%%%%%%%%%%%%%%%%%%%%%%%%%%%%%%%%%%%% 
\maketitle
  
\vspace{0.50cm}
PACS number(s): {21.30.-x, 21.45.Ff, 21.65.-f, 21.65.Ef} 
\vspace{0.50cm}

%%%%%%%%%%%%%%%%%%%%%%%%%%%%%%%%%%%%%%%%%%%%%%%%%%%%%%%%%%%%%%%%%%%%%
\section{Introduction}
%%%%%%%%%%%%%%%%%%%%%%%%%%%%%%%%%%%%%%%%%%%%%%%%%%%%%%%%%%%%%%%%%%%%%
The important role played by the three-nucleon forces (TNFs) has been widely pointed out 
both in finite nuclei and nuclear matter calculations 
(see {\it e.g.}~\cite{kalan12,piper01,epel09,hebeler,holt,mach11,hebeler12,hammer13,hebeler13,tews13,kru13,carbone13,carbone13b,Eks13,zuo14,holt14,gez14,cor13,dri14,rog14} 
and references therein quoted). 
First indications for the inclusion of a TNF in the nuclear Hamiltonian arose from the discrepancy between the results of the $^3$H binding energy using different nucleon-nucleon (NN) 
potentials and its experimental value. For example using high precision NN potentials, able to fit 
NN scattering data up to an energy of $350$ MeV with a $\chi^2$ per datum close to $1$,
the $^3$H, $^3$He and $^4$He binding energies are under-predicted by about $1$ and $4$ MeV 
in the case of the three- or four-nucleon systems respectively~\cite{kievsky2008}. 
A commonly accepted solution to this problem has been the introduction of TNF 
that could bridge the gap between the calculated binding energy \cite{fbs,fgp} based on 
two-body interactions and the experimental binding energies. 
The origin of such a TNF lies in the fact that nucleons are treated as point like 
particles disregarding therefore for their internal quark structure. The TNF emerges 
as a residual {\it tidal} force. 

In nuclear matter calculations based on non-relativistic many-body approaches similar problems arise.  
In such calculations, when only a two-body NN potential is used, symmetric nuclear matter (SMN) results 
over-bound and its empirical saturation point 
$\rho_{0} = 0.16~{\rm fm}^{-3}$,  $E/A|_{\rho_0} = -16$~MeV   
cannot be reproduced. 
As in the case of few-nucleon systems, also for the nuclear matter case TNFs are considered as the missing 
physical effect of the whole picture. 
In addition, TNFs are likely crucial in the case of dense $\beta$-stable nuclear matter to obtain a stiff 
equation of state (EOS) \cite{bbb97,li-hjs08,chamel11} compatible with the measured masses,  
$M = 1.97 \pm 0.04 \, M_\odot$ \cite{demo10} and $M = 2.01 \pm 0.04 \, M_\odot$ \cite{anto13} 
of the neutron stars in PSR~J1614-2230 and PSR~J0348+0432 respectively.    

In relativistic microscopic approaches, such as the Dirac-Brueckner-Hartree-Fock one, the importance of 
three-nucleon interaction is diminished \cite{dbhf}. 
In this approach TNFs are partially included by means of nucleon-antinucleon virtual excitations in the scalar $\sigma$-meson exchange process due
to the dressed Dirac spinor in the nuclear medium. 

Although systems as nuclear matter and finite nuclei deal with the same interactions, the numerical 
calculations of  the properties of these systems are commonly performed using different approaches 
and numerical techniques. 
Moreover, different parametrizations of the same TNF are most of the time present in literature 
according to the system treated: finite nuclei or nuclear matter. 
In fact, the TNF parameters can be fixed to reproduce the properties of few-nucleon ($A = 3,4$) systems, 
or the empirical saturation point of nuclear matter.     
In this paper we analyze the differences existing in the sector of three-nucleon interaction between 
these two areas of nuclear physics.  We would like to see if the need to use different TNF 
parametrization is the consequence of a restricted search in the relative strength of some TNF terms 
or if it is a more fundamental problem.  
For example in Ref.~\cite{pisa} it has been shown that in order to simultaneously describe 
the $^3$H, $^3$He and $^4$He binding energies and the neutron-deuteron (n-d) doublet scattering length 
it is necessary to modify some of the strengths present in the TNF.  
Along this line we investigate the possibility to find a paramatrization of the TNF 
suitable both for finite nuclei and many-body calculations.         

The paper is organized as follows: in the second section we present the three-nucleon force models used 
in this work; in the third section we briefly review the many-body Brueckner-Hartree-Fock (BHF) 
approach and we discuss how to include a TNF in this formalism; 
finally, the fourth section is devoted to show the results of our calculations and to outline the main 
conclusions of this study.

%%%%%%%%%%%%%%%%%%%%%%%%%%%%%%%%%%%%%%%%%%%%%%%%%%%%%%%%%
\section{Three nucleon forces}
%%%%%%%%%%%%%%%%%%%%%%%%%%%%%%%%%%%%%%%%%%%%%%%%%%%%%%%%%
The TNFs that we considered in this work are the new Tucson-Melbourne 
potential \cite{TM'} (hereafter TM') and the three-nucleon potential based on chiral perturbation theory 
calculated at next-to-next-to-leading order \cite{epelbaum02} in its local form \cite{N2LO} (hereafter N2LOL). 
The TM' potential is a revisited version of the older Tucson-Melbourne potential \cite{TM} readjusted in order to satisfy the chiral symmetry. The final operatorial structure coincides with the one obtained in the old Brasilian 
three-nucleon model \cite{brazil}. 
These potentials, in conjunction with the Argonne V18 NN potential \cite{av18}, have been recently used 
by the Pisa group \cite{pisa} to find a new parametrization able to reproduce simultaneously the binding energies of the $^4$He and $^3$He nuclei and the neutron-deuton scattering length $^2$$a_{nd}$.    
The TM' and the N2LOL potentials can be written in the following way \cite{pisa}:

%%%%%%%%%%%%%%%%%%%%%%%%%%%
\begin{figure}[t]
\begin{center}
%\vspace{1cm}
\includegraphics[scale=0.5,angle=0]{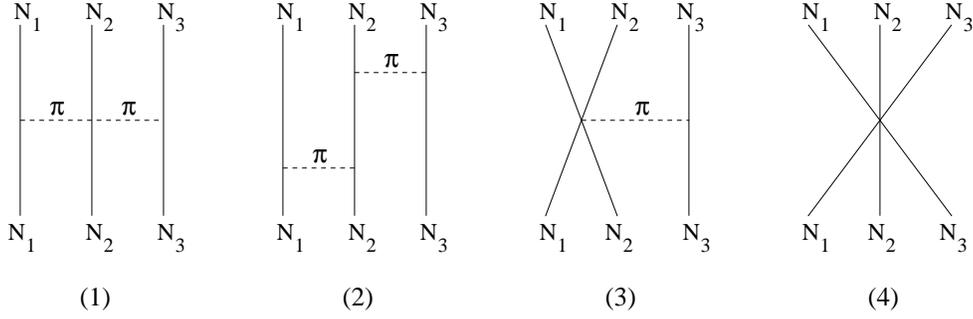}
\caption{Diagrams representing the contribution of terms $a$ (diagram 1), $b, d$ (diagram 2), $D$
 (diagram 3) and $E$ (diagram 4) to the generic three-nucleon force $W(1,2,3)$ of Eq.\ (\ref{eq:w123})
}
\label{fig:diagr}
\end{center}
\end{figure}
%%%%%%%%%%%%%%%%%%%%%%%%%%%%%%%%%%

\begin{equation}
W= \sum_{cyc} W(1,2,3)  \;\; ,
\label{eq:wijk}
\end{equation}
where $W(1,2,3)$ is a generic term that can be put in the following form:
\begin{equation}
W(1,2,3)= aW_a(1,2,3)+bW_b(1,2,3)+dW_d(1,2,3)+c_DW_D(1,2,3)+c_EW_E(1,2,3) \; .
\label{eq:w123}
\end{equation}
In Eq.\ (\ref{eq:w123}) each term corresponds to one of the different mechanism shown in Fig.\ \ref{fig:diagr} and has a different operatorial structure. 
The numerical values of the constants $a$, $b$, $d$, 
$c_D$ and $c_E$ appearing in front of each term of Eq.\ (\ref{eq:w123})
are given in Tab.\ \ref{tab:tab3.1} for each model.
The first three terms arise from the exchange of two pions between the three nucleons (diagrams 1 and 2).
The term $a$ comes from $\pi N$ $S$-wave scattering (diagram 1) 
whereas the terms $b$ and $d$, which are the most important, come from $\pi N$ $P$-wave scattering (diagram 2). 
The specific form of these three terms in configuration space is the following: 

\begin{equation}
\begin{aligned}
& W_a(1,2,3) =-W_0(\bm\tau_1\cdot\bm\tau_2)(\bm\sigma_1\cdot \bm r_{31})
           (\bm\sigma_2\cdot \bm r_{23}) y(r_{31})y(r_{23}) \\
& W_b(1,2,3)= W_0 (\bm\tau_1\cdot\bm\tau_2) [(\bm\sigma_1\cdot\bm\sigma_2)
  y(r_{31})y(r_{23})  \\
 &\hspace{2cm} + (\bm\sigma_1\cdot \bm r_{31})
    (\bm\sigma_2\cdot \bm r_{23})(\bm r_{31}\cdot \bm r_{23})
  t(r_{31})t(r_{23}) \\
 & \hspace{2cm} + (\bm\sigma_1\cdot \bm r_{31})(\bm\sigma_2\cdot \bm r_{31})
  t(r_{31})y(r_{23}) \\
 &\hspace{2cm} + (\bm\sigma_1\cdot \bm r_{23})(\bm\sigma_2\cdot \bm r_{23})
  y(r_{31})t(r_{23})] \\
& W_d(1,2,3)=W_0(\bm\tau_3\cdot\bm\tau_1\times\bm\tau_2)  
   [(\bm\sigma_3\cdot \bm\sigma_2\times\bm\sigma_1)y(r_{31})y(r_{23}) \\
  & \hspace{2cm}+ (\bm\sigma_1\cdot \bm r_{31})
    (\bm\sigma_2\cdot \bm r_{23})(\bm\sigma_3\cdot\bm r_{31}\times \bm r_{23})
  t(r_{31})t(r_{23}) \\
 & \hspace{2cm} 
  + (\bm\sigma_1\cdot \bm r_{31})(\bm\sigma_2\cdot \bm r_{31}\times\bm\sigma_3)
  t(r_{31})y(r_{23}) \\
 &\hspace{2cm} + (\bm\sigma_2\cdot \bm r_{23})(\bm\sigma_3\cdot \bm r_{23}\times
  \bm\sigma_1) y(r_{31})t(r_{23})]\;\; ,
\end{aligned}
\end{equation}
%%%%%%%%%%%%%%%%%%%%%%%%%%%%%%%%%%%%%%%%%%%%
Terms $a$, $b$ and $d$ are present in both the TM' and N2LOL models. The functions $y(r)$ and $t(r)$ 
are defined below. $W_0$ denotes the overall strength 
of these three terms and is defined in a different way in the two models. 
In the TM' case $W_0$ reads:  
\begin{equation}
W_0=\left(\frac{g\,m_\pi}{8\pi\, m_N}\right)^2m_\pi^4 \ ,
\label{eq:w0tm}
\end{equation}
where $g=14.06$, and $m_\pi$ and $m_N$ are the pion and nucleon masses, respectively.  
In the N2LOL model $W_0$ is given by: 
\begin{equation}
W_0=\left(\frac{1}{12\pi}\right)^2\left(\frac{m_\pi}{F_\pi}\right)^4g_A^2m_\pi^2 \ ,
\label{eq:w0n2lol}
\end{equation}
with $F_\pi=92.4$ MeV and $g_A=1.29$.

%%%%%%%%%%%%%%%%%%%%%%%%%%%%%%%%%%%%%%%%%%%%%%%%%%%%%%%%%%%%%%%%%%%%%%%%%%
\begin{table}
\begin{center}
\begin{tabular}{l|ccccc}
\hline
\hline
       & $a$  & $b$ & $d$ & $c_D$ & $c_E$  \\
\hline
%UIX    & $0$            & $1$ & $1/4$     & $0$ & $1$  \\
TM'    & $-0.87/m_\pi$  & $-2.58/m_\pi^3$ & $-0.753/m_\pi^3$ & $0$ & $0$       \\
N2LOL  & $c_1m_\pi^2$   & $c_3/2$         & $c_4/4$          & $1$ & $-0.029$  \\
\hline
\hline
\end{tabular}
\caption{Numerical values of the constants $a$, $b$, $d$, $c_D$ and $c_E$ appearing in front of each 
term of Eq.\ (\ref{eq:w123}) for each model. Constants $c_1=-0.00081$ MeV$^{-1}$,
$c_3=-0.0032$ Mev$^{-1}$ and $c_4=-0.0054$ MeV$^{-1}$ are taken from Ref. \cite{entem}.}
\label{tab:tab3.1}
\end{center}
\end{table}
%%%%%%%%%%%%%%%%%%%%%%%%%%%%%%%%%%%%%%%%%%%%%%%%%%%%%%%%%%%%%%%%%%%%%%%%%%
%

%%%%%%%%%%%%%%%%%%%%%%%%%%%%%%%%%%%%%%%%%%%%%%%%%
 \begin{table} 
 %\begin{center}
 %\bigskip
 \begin{ruledtabular}
\begin{tabular}{l|ccccc}
\hline
  &  $b \ [m^{-3}_\pi]$ & $d \ [m^{-3}_\pi]$ & $c_E$ & $\Lambda \ [m_\pi]$   \\               
\hline
TM1' & -8.256 & -4.690 & 1.0 & 4.0     \\
TM2' & -3.870 & -3.375 & 1.6 & 4.8     \\
TM3' & -2.064 & -2.279 & 2.0 & 5.6     \\
%\hline
%\hline
 \end{tabular}
 \end{ruledtabular}
\caption{Three different parametrizations of the TM' three-body force. 
The value $a=-0.87\; m^{-1}_\pi$ has been kept fix in all the three cases. 
See text and Ref.~\cite{pisa} for details.}
\label{tb:tme}
 %\end{center}
 \end{table}
%%%%%%%%%%%%%%%%%%%%%%%%%%%%%%%%%%%%%%%%%%%%%%%%%%%%%

%%%%%%%%%%%%%%%%%%%%%%%%%%%%%%%%%%%%%%%%%%%%%%%%%%%%%
 \begin{table} 
 %\begin{center}
 %\bigskip
 \begin{ruledtabular}
\begin{tabular}{l|ccccc}
\hline
 &  $c_3$ \ [MeV$^{-1}$] & $c_4$ \ [MeV$^{-1}$] & $c_D$ & $c_E$   \\               
\hline
 N2LOL1 & -0.00448   &  -0.001963  &  -0.5 &   0.100 \\
 N2LOL2 & -0.00448   &  -0.002044  &  -1.0 &   0.000 \\
 N2LOL3 & -0.00480   &  -0.002017  &  -1.0 &  -0.030 \\
 N2LOL4 & -0.00544   &  -0.004860  &  -2.0 &  -0.500 \\
%\hline
%\hline
 \end{tabular}
 \end{ruledtabular}
\caption{Four different parametrizations of the N2LOL three-body force. 
See text and Ref.~\cite{pisa} for details. 
The value $c_1=-0.00081$ MeV$^{-1}$ has been kept fix in all the four cases.}
\label{tb:n2lo}
 %\end{center}
 \end{table} 
%%%%%%%%%%%%%%%%%%%%%%%%%%%%%%%%%%%%%%%%%%%%%%%%%%%%%%%%%%

Term $D$ is present only in the N2LOL model, and it provides the contribution
of a two-nucleon contact term with the emission and absorption of a pion (diagram 3). 
Its local form in  configuration space, derived in Ref. \cite{N2LO} reads: 
\begin{eqnarray}
 W_D(1,2,3)&=& W_0^D (\bm\tau_1\cdot\bm\tau_2) \Big[(\bm\sigma_1\cdot\bm\sigma_2)
  [y(r_{31})Z_0(r_{23})+Z_0(r_{31})y(r_{23})]  \nonumber \\
 &+& (\bm\sigma_1\cdot \bm r_{31})(\bm\sigma_2\cdot \bm r_{31})
  t(r_{31})Z_0(r_{23}) \nonumber \\
 &+& (\bm\sigma_1\cdot \bm r_{23})(\bm\sigma_2\cdot \bm r_{23})
  Z_0(r_{31})t(r_{23})\Big] \;,
\label{eq:wd2}
\end{eqnarray}
where the constant $W_0^D$ is defined as
\begin{equation}
W_0^D=\left(\frac{1}{12\pi}\right)^2\left(\frac{m_\pi}{F_\pi}\right)^4
       \left(\frac{m_\pi}{\Lambda_x}\right)
\frac{g_Am_\pi}{8} \ ,
\label{eq:w0Dn2lol}
\end{equation}
being $\Lambda_x$ the chiral symmetry breaking scale with a value of $700$ MeV. 

Term $E$ is also present only in the N2LOL model but not in the TM' one. 
For the N2LOL model, term $E$ gives the contribution of a   
three-nucleon contact term (diagram 4).
It reads:
\begin{equation}
W_E^{N2LOL}(1,2,3) = W_0^E(\bm\tau_1\cdot\bm\tau_2) Z_0(r_{31})Z_0(r_{23})  \ ,
\label{eq:wen2lol}
\end{equation}
where $W_0^E$ is
\begin{equation}
W_0^E=\left(\frac{1}{12\pi}\right)^2\left(\frac{m_\pi}{F_\pi}\right)^4
       \left(\frac{m_\pi}{\Lambda_x}\right)m_\pi \ .
\label{eq:w0En2lol}
\end{equation}
Although, as said before, the term $E$ is not present in the original TM' model, 
recently in Ref.~\cite{pisa} this model has been extended  by introducing a term similar to that 
of Eq.\ (\ref{eq:wen2lol}):  
\begin{equation}
W_E^{TM'}(1,2,3)=W_0^E Z_0(r_{31})Z_0(r_{23})\ ,
\label{eq:wEtm}
\end{equation}
where for simplicity the isospin dependence has been omitted. We will refer to this 
modification of the original TM' potential also as TM' and it is the one that we will use 
in all the calculations presented in this work. 
Note that for this extended TM' force \cite{pisa} 
the constant $c_E$ is different from zero (see Tab.\ \ref{tb:tme}) and not equal 
to zero (see Tab.\ \ref{tab:tab3.1}).  

The radial dependence of the five terms is encoded in the functions $y(r)$, $t(r)$ and $Z_0(r)$. 
For the TM' and N2LOL models, the functions $y(r)$ and $t(r)$ are:
\begin{equation}
y(r)=\eta_0 \frac{f_0'(r)}{r}
, \,\,\,\, t(r)=\frac{y'(r)}{r} \;,
\label{eq:yttmn2lol}
\end{equation}
where the prime symbol in $f_0^{\prime}(r)$ and $y^{\prime}(r)$ denotes the derivative with respect to $r$, 
the factor $\eta_0$ is equal to $1/3$ for the TM' model and to $1$ for the N2LOL model. 
The function $f_0(r)$ is given by: 
\begin{equation}
f_0(r)=\frac{12\pi}{m_\pi^3}\frac{1}{2\pi^2}
 \int_0^\infty dq q^2 \frac{j_0(qr)}{q^2+m_\pi^2}F_\Lambda(q) \, , 
\label{eq:f0}
\end{equation}
with $j_0(qr) = \sin(qr)/(qr)$.  
The cutoff function $F_\Lambda(q)$ in the TM' model is taken as: 
\begin{equation}
F_\Lambda(q)=\Big[\frac{\Lambda^2-m_\pi^2}{\Lambda^2+q^2}\Big]^2 \;,
\label{eq:cuttm}
\end{equation}
while in the N2LOL model is given by: 
\begin{equation}
F_\Lambda(q)=exp\left(-q^4/\Lambda^4\right) \ . 
\label{eq:cutn2lol}
\end{equation}
$\Lambda$ is a momentum cutoff parameter that fixes the scale of the system in momentum space.
In the N2LOL, it has been set to $\Lambda=500$ MeV, whereas in the TM' model the ratio
$\Lambda/m_\pi$ has been varied in order to describe the $^3$H and $^4$He binding energies
at fixed values of the constants $a$, $b$ and $d$. In literature the TM' potential
has been used in several works (see e.g., Ref. \cite{ref_tm_cutoff}) with typical values 
around $\Lambda= 5\; m_\pi$. 

The function $Z_0(r)$ appearing in Eqs. (\ref{eq:wd2}), (\ref{eq:wen2lol}) and (\ref{eq:wEtm}) 
is defined as: 
\begin{equation}
Z_0(r)=\frac{12\pi}{m_\pi^3}\frac{1}{2\pi^2}
 \int_0^\infty dq q^2 j_0(qr) F_\Lambda(q) \ ,
\label{eq:z0r}
\end{equation}
with $F_\Lambda(q)$ defined in Eq.\ (\ref{eq:cuttm}) for the TM' model and in Eq.\ (\ref{eq:cutn2lol}) 
for the N2LOL one.

%%%%%%%%%%%%%%%%%%%%%%%%%%%%%%%%%%%%%%%%%%%%%%%%
\section{The Brueckner-Hartree-Fock approach}
%%%%%%%%%%%%%%%%%%%%%%%%%%%%%%%%%%%%%%%%%%%%%%%%

The basic ingredient of the BHF approach in nuclear matter \cite{day67,baldo99} is the Brueckner 
reaction matrix $G$ describing the effective interaction between two nucleons in the presence of 
a surrounding medium. 
In the case of asymmetric nuclear matter
%%%%%%%%%%%%%
{\footnote{In the present work we consider spin unpolarized nuclear matter. 
Spin polarazied nuclear matter has been, for example, considered in Ref. \cite{vb02,bomb+06}.}
%%%%%%%%%%%%% 
with neutron density $\rho_n$, proton density $\rho_p$, total nucleon density $\rho = \rho_n + \rho_p$ 
and isospin asymmetry $\beta= (\rho_n-\rho_p)/\rho$ (asymmetry parameter), 
one has different G-matrices describing the nn, pp and np in medium effective interactions.   
They are obtained by solving the well known Bethe--Goldstone equation, written schematically as 

\begin{equation}
G_{\tau_1\tau_2;\tau_3\tau_4}(\omega) = V_{\tau_1\tau_2;\tau_3\tau_4} 
 + \sum_{ij}V_{\tau_1\tau_2;\tau_i\tau_j} \frac{Q_{\tau_i\tau_j}}{\omega-\epsilon_{\tau_i}-\epsilon_{\tau_j} + i\varepsilon}
G_{\tau_i\tau_j;\tau_3\tau_4}(\omega) \;,
\label{bg}
\end{equation}
where $\tau_q$  ($q = 1, 2, i,j, 3, 4$) indicates the isospin projection of the two nucleons in the initial, intermediate and final states, $V$ denotes the bare NN interaction, $Q_{\tau_i\tau_j}$ is the Pauli operator that prevents the intermediate state nucleons $(i, j)$ from being scattered to states below their respective Fermi momenta $k_{F_{\tau}}$ and 
$\omega$, the so-called starting energy, corresponds to the sum of non-relativistic energies of the interacting nucleons. 
The single-particle energy $\epsilon_\tau$ of a nucleon with momentum $k$ and mass $m_\tau$ is given by
\begin{equation}
       \epsilon_{\tau}(k) = \frac{\hbar^2k^2}{2m_{\tau}} + \mbox{Re}[U_{\tau}(k)] \ ,
\label{spe}
\end{equation}
where the single-particle potential $U_{\tau}(k)$ represents the mean field felt by a nucleon due to its
interaction with the other nucleons of the medium. 
In the BHF approximation, $U_{\tau}(k)$ is calculated through the so-called on-energy-shell $G$-matrix, 
and is given by
\begin{equation}
U_{\tau}(k) = \sum_{\tau'} \sum_{k'< k_{F_{\tau'}}} \langle k k'
\mid G_{\tau\tau';\tau\tau'}(\omega=\epsilon_{\tau}(k)+\epsilon_{\tau'}(k')) \mid k k'\rangle_A 
 \;,
\label{spp}
\end{equation}
where the sum runs over all neutron and proton occupied states and the matrix elements are properly antisymmetrized. 
We make use of the so-called continuous choice \cite{jeuk+67,baldo+90,baldo+91} for the single-particle
potential $U_{\tau}(k)$ when solving the Bethe--Goldstone equation. As shown in Refs. \cite{song98,baldo00}, the contribution of the three-hole line diagrams to the energy per particle $E/A$ is minimized in this prescription and a faster convergence 
of the hole-line expansion for $E/A$ is achived \cite{song98,baldo00,baldo90} with respect to the so-called gap choice 
for $U_{\tau}(k)$. 

Once a self-consistent solution of Eqs.\ (\ref{bg})--(\ref{spp}) is achieved, the energy per particle can be calculated as 
\begin{equation}
\frac{E}{A}(\rho,\beta)=\frac{1}{A}\sum_{\tau}\sum_{k < k_{F_{\tau}} }
 \left(\frac{\hbar^2k^2}{2m_{\tau}}+\frac{1}{2} \mbox{Re}[U_{\tau}(k)] \right) \ .
\label{bea}
\end{equation}

%%%%%%%%%%%%%%%%%%%%%%%%%%%%%%%%%%%%%%%%%%%%%%%%%%%%%%%%%%%%%%%%%%
\subsection{Inclusion of three-nucleon forces in the BHF approach}
%%%%%%%%%%%%%%%%%%%%%%%%%%%%%%%%%%%%%%%%%%%%%%%%%%%%%%%%%%%%%%%%%% 
In the microscopic BHF approach the TNFs discussed in the previous section cannot 
be used directly in their original form. This is because it would require the solution of 
a three-body Bethe-Faddeev equation in the nuclear medium and currently this is a task still  
far to be achieved. To avoid this problem an effective density dependent two-body force is built 
starting from the original three-body one by averaging over the coordinates (spatial, spin and isospin) of
one of the three nucleons. The effective NN force due to the NNN one is thus \cite{loiseau,grange89}:

\begin{equation}
W(1,2)=\frac{1}{4} \ Tr_{(\bm\tau_3, \bm\sigma_3)} \int d \rv_3 \ \sum_{cyc}  W(1,2,3) \ n(1,2,3) 
\label{veff}
\end{equation}
In the previous expression $n(1,2,3)$ is the density distribution of the nucleon $3$ in relation 
to the nucleon $1$ at $\rv_1$ and nucleon $2$ at $\rv_2$. 
The function $n(1,2,3)$ represents the effect of the NN correlations and will suppress the contributions from the short-range part of $W(1,2,3)$. 
In the following we adopt an usual choice used in literature \cite{loiseau,grange89}
\begin{equation}
                  n(1,2,3) = \rho \ g^2(1,3) \ g^2(2,3) \; , 
\label{densd}
\end{equation}
where $g(1,3)$ and $g(2,3)$ are the correlation functions between the nucleons $(1,3)$ and $(2,3)$ respectively.    
The latter quantities can be written as $g(1,3) = 1 - \eta(1,3)$, 
where $\eta(1,3)$ is the so-called defect function (and similarly for $g(2,3)\;$). 
Within the BHF approach the defect function should be calculated self-consistently with the 
$G$-matrices (\ref{bg}) and the single particle potentials (\ref{spp}). 
Thus the average effective two-body force (\ref{veff}) should be calculated self-consistently 
and added to the bare NN force at each iterative step of the calculations. 

To simplify the numerical calculations and following \cite{loiseau, grange89}, in the present 
work we use central correlation functions $g(i,j)$ independent on spin and isospin. 
Moreover, it has been shown \cite{BF99,hans} that this central correlation functions, in which are 
included the main contributions of the $^1S_0$ and $^3S_1$ channels, are weakly dependent on the density, 
and can be approximated \cite{loiseau,BF99,hans}  by a Heaviside step function  
$\theta(r_{ij}-r_c)$, with $r_c = 0.6$~fm  in all the considered density range.  
Note that the average procedure has to be performed for each term involved in the cyclic permutation 
in Eq.\ (\ref{eq:wijk}). 
 
In the following we report the expressions we used to perform the reduction of the original TNF 
to the effective density dependent two-body one (\cite{loiseau, grange89}).     

%%%%%%%%%%%%%%%%%%%%%%%%%%%%%%
\begin{figure}
\begin{center}
\vspace{1cm}
\includegraphics[scale=0.8,angle=0]{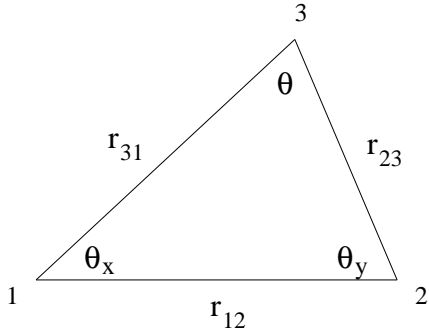}
\caption{ Coordinates of three particle system.}
\label{coord1}
\end{center}
\end{figure}
%%%%%%%%%%%%%%%%%%%%%%%%%%%

For a generic function $F(r_{31},r_{23},r_{12})$ where $r_{31}$, $r_{23}$, $r_{12}$ are the lengths of the three sides of the triangle 
as shown in Fig.\ \ref{coord1} we have: 

%\begin{eqna}
\begin{equation}
\begin{aligned}
& \int d \rv_3 \ \bm\sigma_1\cdot \hat{\bm r}_{31} \bm\sigma_2\cdot \hat{\bm r}_{31} \ F(r_{31},r_{23},r_{12}) \ cos\theta= \\
& \frac{2  \pi}{3  r_{12}} \int_0^{+\infty} d r_{31} \int_{r_{31}+r_{23}}^{|r_{31}-r_{23}|} d r_{23} \ r_{31} \ r_{23} \ F(r_{31},r_{23},r_{12}) \ cos\theta \  
     (\bm\sigma_1\cdot \bm\sigma_2  \ cos\theta+S_{31}(\hat{\bm r}_{31}) \ Q) 
\end{aligned}
%\end{eqna}
\end{equation}

\begin{equation}
%\begin{eqna}
\begin{aligned}
& \int d \rv_3 \ \bm\sigma_1\cdot \hat{\bm r}_{31} \bm\sigma_2\cdot \hat{\bm r}_{31} \ F(r_{31},r_{23},r_{12}) = \\
& \frac{2  \pi}{3  r_{12}}  \int_0^{+\infty} d r_{31} \int_{r_{31}+r_{23}}^{|r_{31}-r_{23}|} d r_{23}    \ r_{31} \ r_{23} \ F(r_{31},r_{23},r_{12})  \  
     (\bm\sigma_1\cdot \bm\sigma_2 \ cos\theta+S_{31}(\hat{\bm r}_{31}) \ Q) 
\end{aligned}
%\end{eqna}
\end{equation}

\begin{equation}
%\begin{eqna}
\begin{aligned}
& \int d \rv_3 \ S_{31}(\hat{\bm r}_{31}) \ F(r_{31},r_{23},r_{12}) = \\
&  S_{12}(\hat{\bm r}_{12})  \ \frac{2  \pi}{3  r_{12}}  \int_0^{+\infty} d r_{31} \int_{r_{31}+r_{23}}^{|r_{31}-r_{23}|} d r_{23}  \ r_{31} \ r_{23} 
\ F(r_{31},r_{23},r_{12})  \   P_2(cos\theta_y) 
\end{aligned}
%\end{eqna}
\end{equation}

Where $cos\theta={\hat{\bm r}_{31}} \cdot {\hat{\bm r}_{23}}$,  $P_2(cos\theta)=\frac{1}{2} (3 \ cos^2\theta-1)$,
 $Q=cos\theta-\frac{3}{2}sin\theta_x sin\theta_y$ (see Fig.\ \ref{coord1} for the definition of angles $\theta$, $\theta_x$ and $\theta_y$) 
and $S_{31}(\hat{\bm r}_{31})=3 \ \bm\sigma_3\cdot \hat{\bm r}_{31} \ \bm\sigma_1\cdot \hat{\bm r}_{31} -\bm\sigma_3\cdot\bm\sigma_1$.

When we consider the term $W(1,2,3)$, the trace operator acting over $\bm\sigma_3$ and $\bm\tau_3$ produces a factor $4$ 
on the terms $W_a$, $W_b$, $W_D$, while makes $W_d$ vanishing due to the traceless property of the $\sigma$ matrices. 
On the other hand in the other two cyclic permutation $W(2,3,1)$ and $W(3,1,2)$ all the previous terms make zero because there is  
always an explicit linear dependence on $\bm\sigma_3$ and $\bm\tau_3$. For the last term $W_E$ we have to discuss separately the TM' and the N2LOL models. 
The N2LOL model has a dependence on $\bm\tau_i \cdot \bm\tau_j$ in the term $W_E$ so also in this case only the permutation $W(1,2,3)$ survives. 
For the TM' model we have instead no isospin dependence in $W_E$ so all the three permutations of $(1,2,3)$  
give contribution to the effective two-body force.
  
Using the above formulas we can perform the two-body reduction of the original three-body force $W$ 
to the effective two-body one. 
The corresponding expressions for the TM' and the N2LOL models can be found in the appendix \ref{appendix}. 
The final effective two-body force is finally added to the bare NN interaction and the energy per particle is 
obtained in BHF approximation as discussed before. 

We want to stress that our average do not take into account some exchange contributions coming from closing a nucleonic fermion line over two different 
nucleons. These contributions are better evaluated starting from the momentum space form of the three-body potential \cite{hebeler}. 
In addition, another possible improvement to the average, is to close the fermionic line considering an interacting propagator \cite{holt}. 
These tasks are beyond the scope of this work and will be considered in the future.

%%%%%%%%%%%%%%%%%%%%%%%%%%%%%%%%%%%%%%%%%%%%%%%%%%%%%%%%%%%%%%%%%%
\section{Results and discussions}
\label{sec:results}
%%%%%%%%%%%%%%%%%%%%%%%%%%%%%%%%%%%%%%%%%%%%%%%%%%%%%%%%%%%%%%%%%%

We now present the results of our calculations of the energy per particle of symmetric nuclear matter (SNM) 
and pure neutron matter (PNM) using the AV18 NN potential supplemented with the TM' or N2LOL 
three-nucleon force. 
Making the usual angular average of the Pauli operator and of the energy denominator \cite{baldo+91,gra87}, 
the Bethe--Goldstone equation (\ref{bg}) can be expanded in partial waves. 
In all the calculations performed in this work, we have considered partial wave contributions up to a total 
two-body angular momentum $J_{max} = 9$.  

Following Ref.~\cite{pisa}, we consider the three parametrizations for the TM' model reported in Tab.\ \ref{tb:tme} 
(hereafter called TM1', TM2' and TM3'), and the four parametrizations of the 
N2LOL model reported in Tab.\ \ref{tb:n2lo} (hereafter called N2LOL1, N2LOL2, N2LOL3 and N2LOL4). 
 
In Fig.\ \ref{snm_TM'_N2LOL} we show the energy per particle $E/A$ of symmetric nuclear matter.  
The green double-dash-dotted line, in both panels, represents the energy per particle with no three-body 
force contribution. The resulting saturation point is 
$\rho_0 = 0.23~{\rm fm}^{-3}$, $E/A|_0 = -16.43~{\rm MeV}$,
to be compared with the empirical saturation point of nuclear matter 
$\rho_{0} = 0.16 \pm 0.01~{\rm fm}^{-3}$, $E/A|_{\rho_0} = -16.0 \pm 1.0~{\rm MeV}$  
(green box in both panels of Fig.\ \ref{snm_TM'_N2LOL}).  
We next introduce the three-body forces of Ref.~\cite{pisa} using the average procedure described in the previous section. 
In the case of the TM' model (left panel) the three-body force produces a sizeable repulsive effect 
({\it i.e.} $E/A$ increases  with respect to the case with no TNF) in all the considered density range,  
and shifts the calculated saturation point (see Tab.\ \ref{satTM}) to a density lower than the empirical one.    
At the empirical saturation density $E/A$ increases by 
$\Delta E = 4.9~{\rm MeV}$ ($6.8~{\rm MeV}$) in the case of the TM1' (TM3') interaction. 
At twice the empirical saturation density, {\it i.e.} $\rho = 0.32~{\rm fm}^{-3}$,  
$\Delta E = 20.9~{\rm MeV}$ ($28.1~{\rm MeV}$) in the case of the TM1' (TM3') interaction. 

The outcome is notably different in the case of the N2LOL (right panel) three-body forces. 
In this case the TNF produces a decrease of $E/A$ in all the considered density range.
At the empirical saturation density (at twice the empirical saturation density) $E/A$ decreases by 
$\Delta E = -1.4~{\rm MeV}$  ($-2.9~{\rm MeV}$) in the case of the N2LOL1 interaction. 
The contrasting effect on the energy per particle of SMN of the two TNF models, illustrated in the two panels 
of Fig.\ \ref{snm_TM'_N2LOL}, is mainly due to the different action of the repulsive component on 
the two-body effective force $W(1,2)$ derived from the genuine TNF. 
As discussed before, the N2LOL model has a nontrivial isospin dependence in the repulsive term $W_E$ 
(see Eq.\ (\ref{tbfN2LOL})). 
If the parameter $C_E$ is positive (negative) the final contribution is repulsive (attractive) on channels 
with isospin $T=1$ but is attractive (repulsive) on channels with isospin $T=0$.  
On the other hand in the TM' model there is no isospin dependence on the repulsive part of the three-body force 
(see Eq.\ (\ref{tbfTM})) so $W_E$ gives in all channels a repulsive contribution. 

The value of the saturation density, energy per particle and symmetry energy at the saturation density
are reported in Tabs. \ref{satTM} and \ref{satN2LOL} for the TM' and the N2LOL models respectively. 
Both models fail to reproduce the empirical saturation point of SNM. 
This is not surprising (see {\it e.g.} \cite{wff88}) since in the present nuclear matter calculations 
we used TNF models \cite{pisa} whose parameters have been determined to reproduce  
the properties of light ($A = 3,\; 4$) nuclei and the neutron-deuteron doublet scattering lenght. 
 
%%%%%%%%%%%%%%%%%%%%%%%%%%%%
\begin{figure}[t]
\begin{center}
%\vspace{1cm}
\includegraphics[scale=0.5,angle=0]{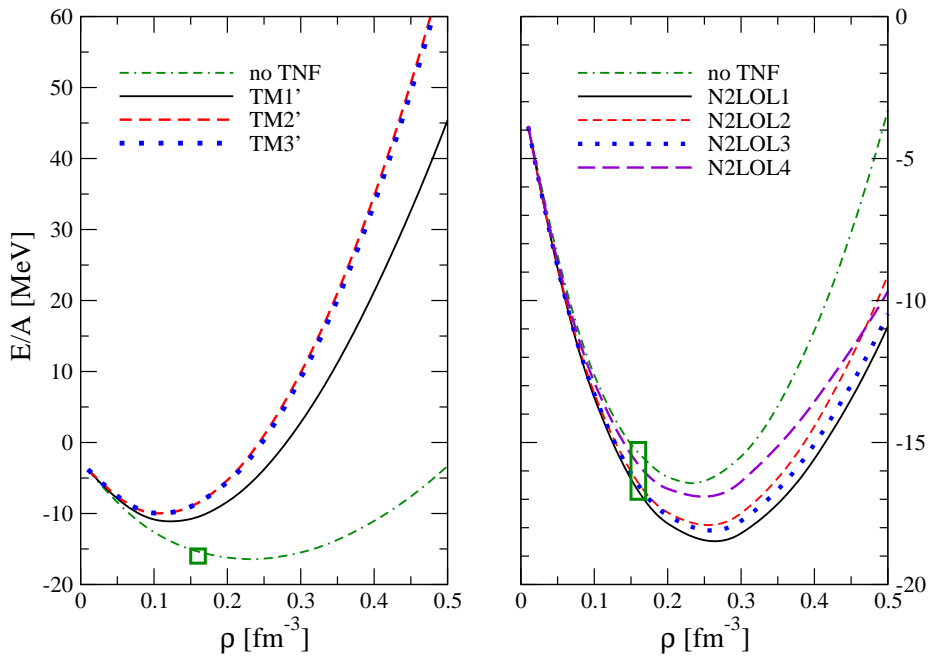}
\caption{(Color on line) Energy per particle $E/A$ of symmetric nuclear matter as a function 
of the nucleonic number density $\rho$ for the three parametrizations of the TM' model (left panel) 
and the four parametrizations of the N2LOL one (right panel) considered in the present work. 
The green double-dash-dotted line, in both panels, represents the energy per particle with no three-body 
force contribution and using the AV18 NN potential.
The empirical saturation point of nuclear matter 
$\rho_{0} = 0.16 \pm  0.01~{\rm fm}^{-3}$, $E/A|_{\rho_0} = -16.0 \pm 1.0~{\rm MeV}$ 
is denoted by the green box in both panels.}
\label{snm_TM'_N2LOL}
\end{center}
\end{figure}
%%%%%%%%%%%%%%%%%%%%%%%%%%

%%%%%%%%%%%%%%%%%%%%%%%%%%%%%%%%%%%%%%%%%%%%%%%%%%%%%%%%%%%%%%%%%%%%%%%%%%  
\begin{table} 
 %\begin{center}
 %\bigskip
 \begin{ruledtabular}
\begin{tabular}{l|cccc}
\hline
          & $\rho_0$ \ (fm$^{-3}$) & $E/A$ \ (MeV) & $E_{sym}$ \ (MeV)   \\               
\hline
 no TNF   & 0.23  & -16.43  & 35.25      \\
  TM1'    & 0.12  & -11.11  & 26.55      \\
  TM2'    & 0.11  &  -9.98  & 22.49      \\
  TM3'    & 0.11  &  -9.96  & 23.05      \\
%\hline
%\hline
 \end{tabular}
 \end{ruledtabular}
\caption{Saturation properties of symmetric nuclear matter for three different parametrizations (first column) of the TM' three-body force. The entry "no TNF" refers to a calculation without three-body force and using 
the AV18 NN potential. The other entries in the table are: the saturation density (second column), the value of energy per particle at saturation (third column) and the value of the symmetry energy at saturation (forth column).}
%\end{center}
\label{satTM}
 \end{table}

%%%%%%%%%%%%%%%%%%%%%%%%%%%%%%%%%%%%%%%%%%%%%%%%%%%%%%%%%%%%%%
 \begin{table} 
 %\begin{center}
 %\bigskip
 \begin{ruledtabular}
\begin{tabular}{l|cccc}
\hline
          & $\rho_0$ \ (fm$^{-3}$) & $E/A$ \ (MeV) & $E_{sym}$ \ (MeV)   \\               
\hline
  N2LOL1    & 0.26 & -18.47  & 42.30      \\
  N2LOL2    & 0.25 & -17.90  & 40.02      \\
  N2LOL3    & 0.26 & -18.09  & 41.06      \\
  N2LOL4    & 0.25 & -16.90  & 36.25      \\ 
%\hline
%\hline
 \end{tabular}
 \end{ruledtabular}
\caption{Saturation properties of symmetric nuclear matter for three different parametrizations of 
the N2LOL three-body force.}
\label{satN2LOL}
 \end{table}
%%%%%%%%%%%%%%%%%%%%%%%%%%%%%%%%%%%%%%%%%%%%%%%%%%%%%%%%%%%%%%%%%%%%%%%%%%

%%%%%%%%%%%%%%%%%%%%%%%%%%%%%%%%%%%%%%%
\begin{figure}[t]
\begin{center}
\vspace{1cm}
\includegraphics[scale=0.5,angle=0]{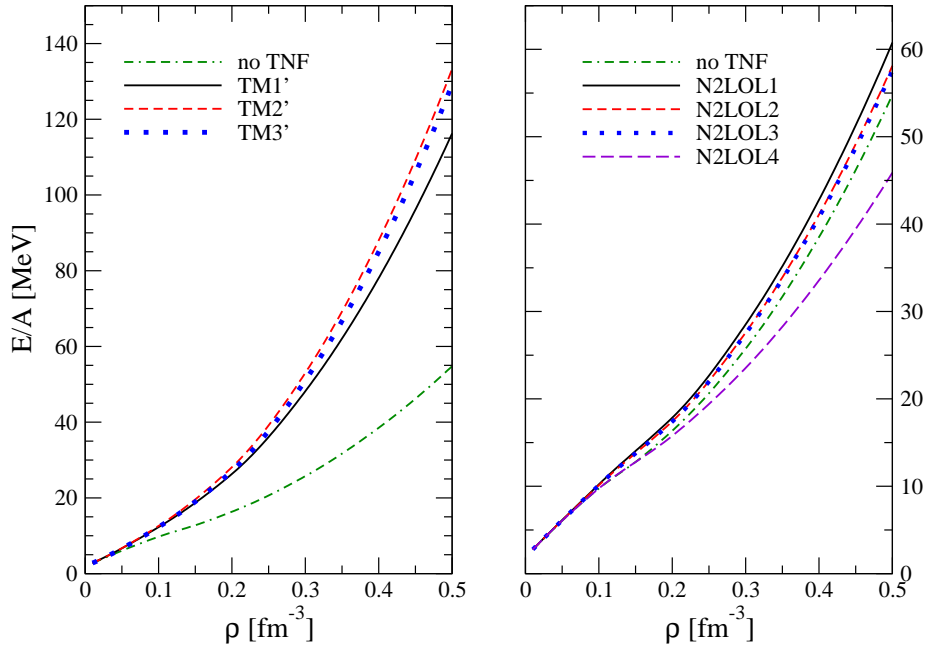}
\caption{
Energy per particle $E/A$ of pure neutron matter as a function of the nucleonic number density $\rho$ 
for the three parametrizations of the TM' model (left panel) and the four parametrizations of 
the N2LOL one (right panel) considered in the present work. 
The green double-dash-dotted line, in both panels, represents the energy per particle with no three-body 
force contribution and using the AV18 NN potential.
}
\label{pnm_TM'_N2LOL}
\end{center}
\end{figure}
%%%%%%%%%%%%%%%%%%%%%%%%%%%%%%%%%%%%%%%%%%%%%

In Fig.\ \ref{pnm_TM'_N2LOL} we plot our results for the energy per particle of pure neutron matter.  
The green double-dash-dotted line, in both panels, represents $E/A$ when the TNF is not included. 
In the case of the TM' model (left panel) the TNF produces a sizeable repulsive effect, in all the 
considered density range, as compared to the case with no TNF.  
For example, at the empirical saturation density (at twice the empirical saturation density) 
$E/A$ increases by $\Delta E = 6.7~{\rm MeV}$ ($25.5~{\rm MeV}$) in the case of the TM1' interaction. 
The effect of TNF on the energy per particle of PNM is less pronounced in the case of the N2LOL model 
(right panel). In the particular case of the N2LOL4 parametrization, TNFs make PNM 
softer with respect to the case where TNFs are not included.   
Notice that in the case of pure neutron matter, we have just the contribution of the $T=1$ isospin channel 
and therefore, also for the N2LOL model one has a pure term that provides repulsion. 
Nonetheless the strength of term associated to repulsion in the TM' model is stronger than  
the corresponding one for the N2LOL and consequently a stiffer neutron matter equation of state 
is obtained. 
%%%%%%%%%%%%%%%%%%%%%%%%%%%
\begin{figure}[t]
\begin{center}
%\vspace{1cm}
\includegraphics[scale=0.5,angle=0]{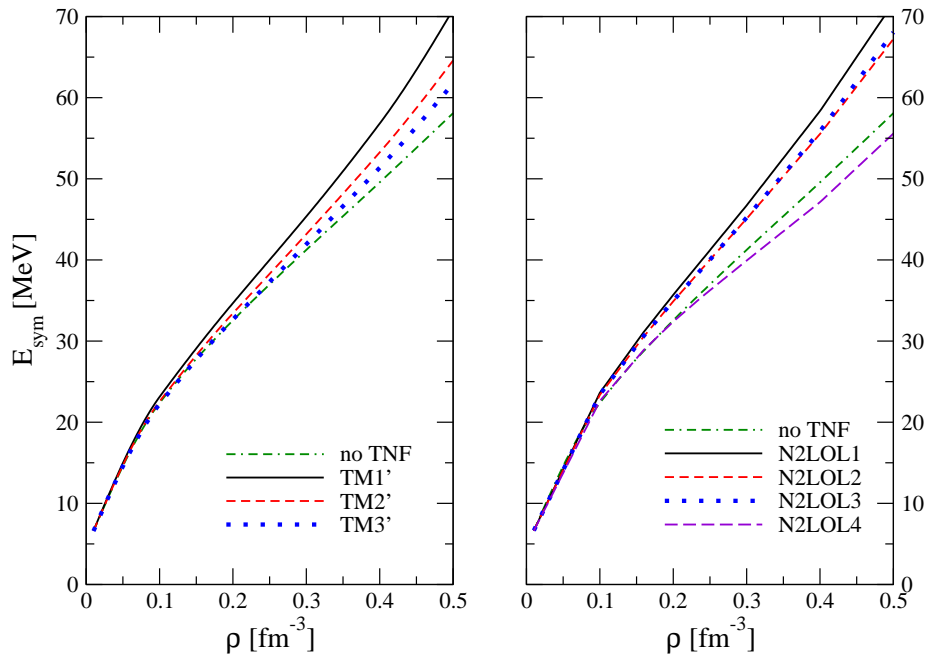}
\caption{Symmetry energy as a function of the nucleonic number density $\rho$ for the three parametrizations 
of the TM' model (left panel) and the four parametrizations of the N2LOL one (right panel) considered 
in the present work. 
The green double-dash-dotted line, in both panels, represents the symmetry energy with no three-body 
force contribution and using the AV18 NN potential.
}
\label{esym_TM'_N2LOL}
\end{center}
\end{figure}
%%%%%%%%%%%%%%%%%%%%%%%%%%%%%%%%%%

The nuclear symmetry energy is defined by 
\begin{equation}
         E_{sym}(\rho) = \frac{1}{2} \frac{\partial^2 E/A}{\partial \beta^2}\Big|_{\beta=0}   \, ,
\label{esym}
\end{equation}
where $E/A$, the energy per particle of asymmetric nuclear matter, is calculated using Eq.\ (\ref{bea}) 
within the BHF approximation.  

It has been numerically demonstrated by the authors of Ref. \cite{bl91} and afterwards confirmed by various microscopic calculations \cite{huber95,bordbar98,kuo98,zbl99,zuo2002b,gad07,vid09} adopting different 
realistic NN interactions, that the energy per particle of asymmetric nuclear matter can be accurately 
reproduced by the following relation:  
\begin{equation}
      \frac{E}{A}(\rho, \beta) = \frac{E}{A}(\rho, 0) + E_{sym} (\rho) \beta^2 \,.
\label{eq:parab}
\end{equation}
Thus, in good approximation, the symmetry energy can be expressed in terms of the difference of the energy per particle between symmetric ($\beta = 0$) and pure neutron matter ($\beta = 1$):
\begin{equation}
     E_{sym}(\rho) = \frac{E}{A}(\rho,1) - \frac{E}{A}(\rho,0) \,.
\label{eq:Esym2}
\end{equation}

The symmetry energy for the TM' and the N2LOL models is shown as function of the density on the left and right panels of Fig.\ \ref{esym_TM'_N2LOL} respectively. 
The green double-dash-dotted line, in both panels, represents the symmetry energy with no three-body 
force contribution and using the AV18 NN potential.  
In Tabs.\ \ref{satTM} and \ref{satN2LOL} we have reported the values of the symmetry energy 
for the two TNF models at their respective calculated saturation points $\rho_0$ 
(second column in Tabs.\ \ref{satTM} and \ref{satN2LOL}). 

To compare our results with the value of the symmetry energy extracted from various nuclear experimental 
data \cite{tsang12,latt13}, we report in Tab.\ \ref{esym_nm} $E_{sym}$  calculated at the 
empirical saturation density $\rho_{0} = 0.16~{\rm fm}^{-3}$ for the different TNF models considered in this work. 
In the same table, we also report the so called slope parameter 
\begin{equation}
       L = 3 \rho_{0} \frac{\partial E_{sym}(\rho)}{\partial \rho}\Big|_{\rho_{0}} \,.
\label{slope}
\end{equation}
As we can see (Tab.\ \ref{esym_nm}) our calculated $E_{sym}$ and $L$   
lies within the ranges of values extracted from experimental data \cite{latt13}:  
$E_{sym}(\rho_{0}) = 29.0$ -- $32.7$~MeV, and $L = 40.5$ -- $61.9$~MeV.  

%%%%%%%%%%%%%%%%%%%%%%%%%%%%%%%%%%%%%%%%%%%%%%%%%%%%%%%%%%%%%%
 \begin{table} 
 %\begin{center}
 %\bigskip
 \begin{ruledtabular}
\begin{tabular}{l|ccc}
\hline
          & $E_{sym}$ \ (MeV)   & $L$ \ (MeV)  \\               
\hline
\hline
 no TNF   & 28.79   &  51.3      \\
  TM1'    & 30.14   &  53.6      \\
  TM2'    & 29.30   &  50.3      \\
  TM3'    & 28.65   &  48.4      \\
  N2LOL1  & 31.14   &  56.6      \\
  N2LOL2  & 30.59   &  53.7      \\
  N2LOL3  & 30.65   &  54.1      \\
  N2LOL4  & 28.92   &  44.8      \\ 
%\hline
%\hline
 \end{tabular}
 \end{ruledtabular}
\caption{Symmetry energy and slope parameter $L$ at the empirical saturation density 
$\rho_{nm} = 0.16~{\rm fm}^{-3}$ for different TNF models. }
%\end{center}
\label{esym_nm}
 \end{table}
%%%%%%%%%%%%%%%%%%%%%%%%%%%%%%%%%%%%%%%%%%%%%%%%%%%%%%%%%%%%%%%%%%%%%%%%%%

The TNFs of Ref.~\cite{pisa} have been recently employed for SNM and PNM calculations  
in \cite{lovato12} using both the variational and the auxiliary field diffusion Monte Carlo approaches. 
The authors of Ref.~\cite{lovato12} used the Argonne V8' \cite{av8} NN potential which is a simplified version 
of the AV18 potential \cite{av18}, truncated after the linear spin-orbit components and refitted 
to have the same isoscalar part of the AV18 in all the $S$ and $P$ waves, as well as in the $^3D_1$ wave and its coupling to the $^3S_1$ wave. Notice that also in \cite{lovato12} no one of the TNF models 
was able to reproduce the correct saturation point of symmetric nuclear matter.   
A direct comparison of our results with those reported in \cite{lovato12} would be ambiguous and 
inconclusive for the following reasons. 
First of all due to the difference in the two-body used in the calculations: the AV18 in the present paper 
and the AV8' in \cite{lovato12}.  Second, as discussed in \cite{bomb05,baldo12}, the use of 
different many-body approaches affects considerably the results particularly in the case of SNM.  
   
%%%%%%%%%%%%%%%%%%%%%%%%%%%%%%%%%%%%%%%%%%%%%%%%
\section{Conclusions}
\label{sec:conclusions}
%%%%%%%%%%%%%%%%%%%%%%%%%%%%%%%%%%%%%%%%%%%%%%%% 

A new generation of TNF models has been recently proposed in Ref.~\cite{pisa}. 
These new TNFs have been used, in conjuction with the Argonne V18 two-nucleon interaction, 
and their parameters have been determined to simultaneously reproduce the measured binding energies
of $^3$H, $^3$He and $^4$He nuclei as well as the measured n-d doublet scattering length. 
A correct prediction for these physical quantities can be regarded as a severe requisite for a realistic 
nuclear Hamiltonian containing two- and three-nucleon interactions.  
As shown in Ref.~\cite{pisa}, this requirement was not fulfilled by several of the TNF models available 
in the literature.    

In the present work, we have calculated the energy per nucleon of symmetric nuclear matter and 
pure neutron matter within the BHF approach and using the same nuclear Hamiltonian 
as the one used in Ref.~\cite{pisa} ({\it i.e.} without changing the original values of the TNF parameters) 
with the purpose to test this Hamiltonian in a many-body context. 
We found that none of the TNF models given in \cite{pisa} is able to reproduce the empirical saturation point 
of symmetric nuclear matter. 
This outcome concords with the results obtained with the Urbana VII \cite{rocco86} TNF  
when used in few-body nuclei and nuclear matter \cite{wff88} (see also \cite{bbb97}).  
In particular, in the case of the AV18+TM' Hamiltonian, both the calculated saturation density and 
the corresponding binding energy per nucleon ($B/A = -E/A$) are understimated. 
The TM' model for the TNF produces a strong repulsive effect both in SNM and PNM, in all the considered density range. 
In the case of the AV18+N2LOL Hamiltonian, the TNF produces a decrease of $E/A$ 
(an increase of the binding energy per nucleon) 
in all the considered density range. 

The reasons why TNFs fitting few-nucleon systems are not able to reproduce the empirical saturation 
point of SNM can be various.  
First of all in the BHF approach we are still not able to use the genuine three-body force but, 
due to the technical reasons explained above, we are forced to include the TNF contribution 
by performing the average procedure to generate an effective density dependent two-body force.  
In this way some terms get lost in the average and therefore, in a way, we are not dealing with exactly 
the same three-body force employed in finite nuclei calculations. 
An interesting test could be the use of our effective two-nucleon force in finite nuclei calculations. 
So doing, we may understand at a deeper level, the differences between the exact procedure in which we employ the genuine three-nucleon force and our simplified approach that involve an average force. 
We plan to perform this interesting comparison in a future work. 
In addition, the inclusion of the exchange terms in the average or the use of a correlation function explicitly dependent from spin and isospin will be definitely an improvement of our calculations. 
These aspects will be considered in a future work.

\section{Appendix}
\label{appendix}

The explicit expressions for the TM' and the N2LOL models are reported in the following. 
For the TM' model we have: 

\begin{equation}
V^{TM'}(r_{12})=\frac{\rho}{3} \ (\bm\tau_1\cdot\bm\tau_2) \ [ \bm\sigma_1\cdot\bm\sigma_2 \ v_\sigma(r_{12})+ 
S_{12}(\hat{\bm r}_{12}) \ v_t(r_{12})  + v_r(r_{12}) ]  \;.
\label{tbfTM}
\end{equation}

\begin{equation}
\begin{aligned}
& v_\sigma(r_{12})= 
\frac{2 \pi}{r_{12}} \int_0^{+\infty} d r_{31} \int_{r_{31}+r_{23}}^{|r_{31}-r_{23}|} d r_{23} \ r_{31} \ r_{23} \  
[ 
-\frac{a W_0}{3} r_{31} \ r_{23} \ y(r_{31}) \ y(r_{23}) \ cos\theta+ \\
&\hspace{2cm} b \ W_0 \ r_{31} \ r_{23} \ y(r_{31}) \ y(r_{23}) + 
 \frac{b \ W_0}{3} r_{31}^2 \ r_{23}^2 \ cos^2\theta+  
\frac{b \ W_0}{3} ( r^2_{31} \ t(r_{31}) \ y(r_{23}) + r^2_{23} \ t(r_{23}) y(r_{31}) \ )  \ ] \\
&\hspace{2cm} \  g^2(r_{31}) \  g^2(r_{23})   \;.\\
\end{aligned}
\end{equation}

\begin{equation}
\begin{aligned}
& v_t(r_{12}) =\frac{2 \pi}{r_{12}} \int_0^{+\infty} d r_{31} \int_{r_{31}+r_{23}}^{|r_{31}-r_{23}|} d r_{23} \ r_{31} \ r_{23} 
[
-\frac{a \ W_0}{3} r_{31} \ r_{23} \ y(r_{31}) \ y(r_{23}) \ Q+ \\  
&\hspace{2cm} 
\frac{b \ W_0}{3} \ 
( \ 
 r^2_{31} \ t(r_{31}) \ r^2_{23} \ t(r_{23}) \ cos\theta \ Q 
+ P_2(cos\theta_y) \ r^2_{31} \ t(r_{31}) \ y(r_{23}) +     \\
&\hspace{2cm}
  P_2(cos\theta_x) \ r^2_{23} \ t(r_{23}) \ y(r_{31}) \ ) 
] \ g^2(r_{31}) \ g^2(r_{23})  \;.
\end{aligned}
\end{equation} 

\begin{equation}
\begin{aligned}
& v_r(r_{12})=C_E \ W^{E}_{0} \frac{2 \pi}{r_{12}} \int_0^{+\infty} d r_{31} \int_{r_{31}+r_{23}}^{|r_{31}-r_{23}|} d r_{23} \ r_{31} \ r_{23} 
 [ \ z_0(r_{31}) \ z_0(r_{23})+  \\ 
&\hspace{2cm} z_0(r_{31}) \ z_0(r_{12})+z_0(r_{12}) \ z_0(r_{23}) \ ] 
 \ g^2(r_{31}) \  g^2(r_{23}) \;. \\
\end{aligned}
\end{equation}

%%%%%%%%%%%%%%%%%%%%%%%%%%%%%%%%%%%%%%%%%%%%%%%%%%%%%%%%%%%%%%%%%%%%%%%

For the N2LOL model we have: 

\begin{equation}
V^{N2LOL}(r_{12})=\frac{\rho}{3} \ (\bm\tau_1\cdot\bm\tau_2) \ [ \bm\sigma_1\cdot\bm\sigma_2 \ \tilde{v}_{\sigma}(r_{12})+ 
S_{12}(\hat{\bm r}_{12}) \ \tilde{v}_{t}(r_{12}) + \tilde{v}_{r}(r_{12}) ] \;.
\label{tbfN2LOL}
\end{equation}

\begin{equation}
\begin{aligned}
& \tilde{v}_\sigma(r_{12})= 
\frac{2 \pi}{r_{12}} \int_0^{+\infty} d r_{31} \int_{r_{31}+r_{23}}^{|r_{31}-r_{23}|} d r_{23} \ r_{31} \ r_{23} \  
[ 
-\frac{a W_0}{3} r_{31} \ r_{23} \ y(r_{31}) \ y(r_{23}) \ cos\theta+ \\
&\hspace{2cm} b \ W_0 \ r_{31} \ r_{23} \ y(r_{31}) \ y(r_{23}) + 
 \frac{b \ W_0}{3} r_{31}^2 \ r_{23}^2 \ cos^2\theta+  
\frac{b \ W_0}{3} ( r^2_{31} \ t(r_{31}) \ y(r_{23}) + r^2_{23} \ t(r_{23}) y(r_{31}) )+ \\
&\hspace{2cm}\frac{C_D \ W_0^D}{3}  ( \ r^2_{31} \ t(r_{31}) \ z(r_{23}) + r^2_{23} \ t(r_{23}) z(r_{31}) \ )
\ ] \ g^(r_{31}) \  g^2(r_{23})   \,.\\
\end{aligned}
\end{equation}

\begin{equation}
\begin{aligned}
& \tilde{v}_t(r_{12}) =\frac{2 \pi}{r_{12}} \int_0^{+\infty} d r_{31} \int_{r_{31}+r_{23}}^{|r_{31}-r_{23}|} d r_{23} \ r_{31} \ r_{23} 
[
-\frac{a \ W_0}{3} r_{31} \ r_{23} \ y(r_{31}) \ y(r_{23}) \ Q+ \\  
&\hspace{2cm} 
\frac{b \ W_0}{3} \ 
( \ 
 r^2_{31} \ t(r_{31}) \ r^2_{23} \ t(r_{23}) \ cos\theta \ Q 
+ P_2(cos\theta_y) \ r^2_{31} \ t(r_{31}) \ y(r_{23}) +     \\
&\hspace{2cm}
  P_2(cos\theta_x) \ r^2_{23} \ t(r_{23}) \ y(r_{31}) \ ) +  \\
&\hspace{2cm} 
\frac{C_D \ W_0^D}{3} \  ( \ r^2_{31} \ t(r_{31}) \ z(r_{23}) P_2(cos\theta_y) + r^2_{23} \ t(r_{23}) z(r_{31}) P_2(cos\theta_x) \ ) \
] \ g^2(r_{31}) \  g^2(r_{23}) \;. 
\end{aligned}
\end{equation} 

\begin{equation}
\begin{aligned}
& \tilde{v}_r(r_{12})=C_E \ W^{E}_{0} \frac{2 \pi}{r_{12}} \int_0^{+\infty} d r_{31} \int_{r_{31}+r_{23}}^{|r_{31}-r_{23}|} d r_{23} \ r_{31} \ r_{23} 
 [ \ z_0(r_{31}) \ z_0(r_{23}) \ ] \ 
 \ g^2(r_{31}) \  g^2(r_{23}) \;.\\
\end{aligned}
\end{equation} 

%%%%%%%%%%%%%%%%%%%%%%%%%%%%%%%%%%%%%%%%%%%%%%%%%%%%%%%%%%%%
\section*{Acknowledgments}

This work has been partially supported by the project PEst-OE/FIS/UI0405/2014 developed under the inititative QREN financed by the UE/FEDER throught the program
COMPETE-``Programa Operacional Factores de Competitividade'', and by ``NewCompstar'', COST Action MP1304.

%%%%%%%%%%%%%%%%%%%%%%%%%%%%%%%%%%%%%%%%%%%  

%%%%%%%%%%%%%%%%%%%%%%%%%%%%%%%%%%%%%%%%%%%%%%%%%%%%%%%%%%%%%%%%%%%%%

\begin{thebibliography}{200}
%%%%%%%%%%%%%%%%%%%%%%%%%%%%%%%%%%%%%%%%%%%

\bibitem{kalan12} N. Kalantar-Nayestanaki, E. Epelbaum, J. S. Messchendorp and A. Nogga, 
                  Rep. Prog. Phys. {\bf 75}, 016301 (2012)

\bibitem{piper01}  S. C. Pieper, V. R. Pandharipande, R. B. Wiringa and J. Carlson, Phys. Rev. C {\bf 64}, 
014001 (2001)
%%
\bibitem{epel09} E. Epelbaum, H.-W. Hammer and U.-G. Meissner, Rev. Mod. Phys. {\bf 81}, 1773 (2009).

\bibitem{hebeler}  K. Hebeler and A. Schwenk, Phys. Rev. C {\bf 82}, 014314 (2010); 
K. Hebeler, S. K. Bogner, R. J. Furnstahl, A. Nogga and A. Schwenk, Phys. Rev. C {\bf 83}, 031301 (2011). 

\bibitem{holt} J. W. Holt, N. Kaiser and W. Weise, Phys. Rev. C {\bf 81}, 024002 (2010).   

\bibitem{mach11} R. Machleidt and D. R. Entem, Phys. Rep. {\bf 503}, 1 (2011). 

\bibitem{hebeler12} K. Hebeler, Phys. Rev C {\bf 85}, 021002(R) (2012).

\bibitem{hammer13} H.-W. Hammer, A. Nogga and A. Schwenk, Rev. Mod. Phys. {\bf 85} 197 (2013).

\bibitem{hebeler13} K. Hebeler and R. J. Furnstahl, Phys. Rev. C {\bf 87}, 031302(R) (2013).

\bibitem{tews13} I. Tews, T. Kr\"{u}ger, K. Hebeler and A. Schwenk, Phys. Rev. Lett. {\bf 110}, 032504 (2013).

\bibitem{kru13} T. Kr\"{u}ger, I. Tews, K. Hebeler and A. Schwenk, Phys. Rev. C 88, 025802 (2013).

\bibitem{carbone13} A. Carbone, A. Polls and A. Rios Phys. Rev. C {\bf 88}, 044302 (2013).

\bibitem{carbone13b} A. Carbone, A. Cipollone, C. Barbieri, A. Rios and A. Polls, Phys. Rev. C {\bf 88}, 054326 (2013).

\bibitem{Eks13} A. Ekstr\"{o}m, G. Baardsen, C. Forss\'{e}n, G. Hagen, M. Hjorth-Jensen, G. R. Jansen, R. Machleidt, W. Nazarewicz, T. Papenbrocl, J. Sarich and S. M. Wild, Phys. Rev. Lett. {\bf 110}, 192502 (2013)

\bibitem{zuo14}  W. Zuo, I. Bombaci and U. Lombardo, Eur. Phys. J. A {\bf 50}, 12 (2014)

\bibitem{holt14} J. D. Holt, J. Men\'{e}ndez, J. Simonis and A. Schwenk, Phys. Rev. C {\bf 90}, 024312 (2014).

\bibitem{gez14} A. Gezerlis, I. Tews E. Epelbaum, M. Freunek, S. Gandolfi, K. Hebeler, A. Nogga and A. Schwenk,
	Phys. Rev. C {\bf 90}, 054323 (2014).
%%%
\bibitem{cor13} L. Coraggio, J. W. Holt, N. Itaco, R. Machleidt, and F. Sammaruca, Phys. Rev. C {\bf 87}, 014322 (2013); 
	L. Coraggio, J. W. Holt, N. Itaco, R. Machleidt, L. E. Marcucci, and F. Sammaruca, 
	Phys. Rev. C {\bf 89}, 044321 (2014). 
\bibitem{dri14} C. Drischler, V. Som\'a, and A. Schwenk, Phys. Rev. C {\bf 89}, 025806 (2013). 
\bibitem{rog14} A. Roggero, A. Mukherjee, and F. Pederiva, Phys. Rev. Lett. {\bf 112}, 221103 (2014).
%%%


%%%
%

\bibitem{kievsky2008} A. Kievsky, S. Rosati, M. Viviani, L.E. Marcucci, and
              L. Girlanda, J. Phys. G {\bf 35}, 063101 (2008)

\bibitem{fbs} T. Sasakawa and S. Ishikawa. Few-Body Syst. {\bf 1},3 (1986)

\bibitem{fgp} J. L. Friar, B. F. Gibson and G. L. Payne Phys. Rev. C {\bf 37}, 2869 (1988)

\bibitem{bbb97}   M. Baldo, I. Bombaci and G. F. Burgio, Astron. and Astrophys. {\bf 328}, 274 (1997)

\bibitem{li-hjs08} Z. H. Li and H.-J. Schulze, Phys. Rev. C {\bf 78}, 028801 (2008)

\bibitem{chamel11} N. Chamel, A. F. Fantina, J. M. Paearson and S. Goriely, 
                   Phys. Rev. C {\bf 84}, 062802(R) (2011)

\bibitem{demo10} P. Demorest, T. Pennucci, S. Ransom, M. Roberts, and J. Hessels, 
                           Nature {\bf 467}  (2010) 1081. 

\bibitem{anto13} J. Antoniadis et al., Science {\bf 340} (2013) 1233232. 

\bibitem{dbhf} R. Machleidt, K. Holinde and Ch. Elster Phys. Rep. {\bf 149} 1 (1987); 
               D. Alonso and F. Sammarruca Phys. Rev. C {\bf 67} 054301 (2003); 
               E. Van Dalen and F. C. Faessler Nucl. Phys. A {\bf 744} 227 (2004). 

\bibitem{pisa}  A. Kievsky, M. Viviani and L. Girlanda and E. Marcucci Phys. Rev. C {\bf 81} 044003 (2010)

\bibitem{TM'} S.A. Coon  and H.K. Han, Few-Body Syst. {\bf 30}, 131 (2001)

\bibitem{epelbaum02} E. Epelbaum {\sl et al.}, Phys. Rev. C {\bf 66},
 064001 (2002).

\bibitem{N2LO}   P. Navratil, Few-Body Syst. {\bf 41}, 117 (2007).

\bibitem{TM}    S.A. Coon and W. Gl\"ockle, Phys. Rev. C {\bf 23}, 1790 (1981)

\bibitem{brazil} H.T. Coelho, T.K. Das, and M.R. Robilotta, Phys. Rev. C {\bf 28}, 1812 (1983);
	         M.R. Robilotta and H.T. Coelho, Nucl. Phys. A {\bf 460}, 645 (1986)

\bibitem{av18}    R. B. Wiringa, V. G. J. Stoks and R. Schiavilla, Phys. Rev. C {\bf 51}, 38 (1995)

\bibitem{entem} D.R.\ Entem and R.\ Machleidt,
                Phys.\ Rev.\ C {\bf 68} 041001(R) (2003).

\bibitem{ref_tm_cutoff} A. C. Hayes, P. Navratil and J. P. Vary, Phys. Rev. Lett. {\bf 91}, 01250 (2003); 
                        P. Navratil and W. E. Ormand, Phys. Rev. C {\bf 68}, 034305 (2003). 


%%%%% BHF
\bibitem{day67}  B. D. Day, Rev. Mod. Phys. {\bf 39} 719 (1967)

\bibitem{baldo99}  M. Baldo, in Nuclear Methods and the Nuclear Equation of State, 
         edited by M. Baldo, International Review of Nuclear Physics Vol. 8 
         (World Scientific, Singapore, 1999), p. 1.

\bibitem{jeuk+67}  J. P. Jeukenne, A. Lejeunne, C. Mahaux, Phys. Rep. {\bf 25} 83 (1976)

\bibitem{baldo+90} M. Baldo, I. Bombaci, G. Giansiracusa, U. Lombardo, C. Mahaux and R. Sartor,  
                      Phys. Rev. C {\bf 41} 1748 (1990)

\bibitem{baldo+91} M. Baldo, I. Bombaci, L. S. Ferreira,  G. Giansiracusa, and U. Lombardo, 
                      Phys. Rev. C {\bf 43} 2605 (1991)

\bibitem{vb02}   I. Vida\~na and I. Bombaci, Phys. Rev. C {\bf 66} 045801 (2002)

\bibitem{bomb+06} I. Bombaci, A. Polls, A. Ramos, A. Rios and I. Vida\~na, 
                   Phys. Lett. B {\bf 632} 638 (2006)

\bibitem{song98} H. Q. Song, M. Baldo, G. Giansiracusa and U. Lombardo, Phys. Rev. Lett. {\bf 81}, 1584 (1998).

\bibitem{baldo00} M. Baldo, G. Giansiracusa, U. Lombardo and H. Q. Song, Phys. Lett. B {\bf 473}, 1 (2000).

\bibitem{baldo90} M. Baldo, I. Bombaci, G. Giansiracusa, and U. Lombardo, 
                  J. Phys. G: Nucl. Part. Phys. {\bf 16}, L263 (1990).

\bibitem{loiseau}
 B. A. Loiseau, Y. Nogami and C. K. Ross Nucl. Phys. {\bf A401} 601 (1971). 

\bibitem{grange89}
 P. Grang\'e, A. Lejeunne, B. Martzolff, and J.-F. Mathiot, Phys. Rev.C {\bf 40} 1040 (1989). 

\bibitem{BF99} M. Baldo and L. S. Ferreira, Phys. Rev. C {\bf 59} 682 (1999).

\bibitem{hans} X. R. Zhou, G. F. Burgio, U. Lombardo, H.-J. Schulze, and W. Zuo, 
                Phys Rev. C {\bf 69} 018801 (2004).

\bibitem{gra87}  P. Grang\'e, J. Cugnon, and A. Lejeune, Nucl. Phys. A {\bf 473}, 365 (1987).   

\bibitem{wff88}R. B. Wiringa, V. Fiks, and A. Fabrocini, Phys. Rev. C {\bf 38}, 1010 (1988).

%%% ref. Esym %%%%
\bibitem{bl91} I. Bombaci, U. Lombardo, Phys. Rev. C \textbf{44}, (1991) 1892.
\bibitem{huber95} H. Huber, F. Weber, and M. K. Wiegel, Phys. Rev. C \textbf{51}, (1995) 1790.
\bibitem{bordbar98} G. H. Bordbar and M. Modarres, Phys. Rev. C \textbf{57}, (1998) 714.
\bibitem{kuo98} C. H. Lee, T. T. S. Kuo, G. Q. Li and G. E. Brown, Phys. Rev. C \textbf{57} (1998) 3488
\bibitem{zbl99} W. Zuo, I. Bombaci, and U. Lombardo, Phys. Rev. C \textbf{60}, (1999) 024605.
\bibitem{zuo2002b} W. Zuo, A. Lejeune, U. Lombardo and J. F. Mathiot,
                    Eur. Phys. J. A \textbf{14}, (2002) 469
\bibitem{gad07} Kh. Gad and Kh. S. A. Hassaneen, Nucl. Phys. A \textbf{793}, (2007) 67
\bibitem{vid09} I. Vida\~na, C. Providencia, A. Polls, and A. Rios, Phys. Rev. C \textbf{80} 045806 (2009) 
%%%%%%%%%%%%%%%%%%

\bibitem{tsang12}  M. B. Tsang, J. R. Stone, F. Camera, P. Danielewicz, S. Gandolfi, K. Hebeler, 
                   C. J. Horowitz, Jenny Lee, W. G. Lynch, Z. Kohley, R. Lemmon, P. Moller, 
                   T. Murakami, S. Riordan, X. Roca-Maza, F. Sammarruca, A. W. Steiner, I. Vida\~na 
                   and S. J. Yennello, Phys. Rev. C {\bf 86}, 015803 (2012)

\bibitem{latt13} B. A. Li, A. Ramos, G. Verde and I. Vida\~na, Eds. 
	{\it Topical issue on Nuclear Symmetry Energy}, Eur. Phys. J. A {\bf 50} issue 2 (2014).

\bibitem{lovato12} A. Lovato, O. Benhar, S. Fantoni, and K. E. Schmidt Phys. Rev. C {\bf 85} 024003 (2012).

\bibitem{av8}   B. S. Pudliner, V. R. Pandharipande, J. Carlson, S. C. Piper and R. B. Wiringa, 
                Phys. Rev. C {\bf 56}, 1720 (1997)

\bibitem{bomb05}  I. Bombaci, A. Fabrocini, A. Polls and I. Vida\~na, Phys. Lett B {\bf 609}, 232 (2005).


\bibitem{baldo12} M. Baldo, A. Polls, A. Rios, H.-J. Schulze, and I. Vida\~na,
                  Phys. Rev. C {\bf 86} 064001 (2012). 

\bibitem{rocco86}  R. Schiavilla, V. R. Pandharipande and R. B. Wiringa, Nucl. Phys. A {\bf 449}, 219 (1986)

\end{thebibliography}
\end{document}